\documentclass[aps,pre,onecolumn,superscriptaddress,groupedaddress]{revtex4}  

\usepackage{amsmath}

\usepackage{verbatim}
\usepackage{graphicx}
\usepackage{hyperref}
\usepackage{mathtools}
\usepackage{bm}
\usepackage{lipsum}
\begin{document}
	\title{  Limitations on concentration measurements and    gradient discerning times in cellular systems }
\author{Vaibhav Wasnik}
\email{wasnik@iitgoa.ac.in}

\affiliation{Indian Institute of Technology, Goa}
\begin{abstract}
	 This work reports on two results. At first we revisit      the Berg and Purcell calculation that provides a lower bound to the error in  concentration measurement by cells,  by considering the   realistic case when the cell starts measuring the moment it comes in contact with  the chemoattractants, instead of measuring   after equilibrating with the chemotactic concentration as done in the classic Berg and Purcell paper.    We find that the error in concentration measurement is  still the same   as   evaluated by    Berg and Purcell.   We next derive a lower bound  on measurement  time below which it is not possible for the cell to   discern extra-cellular chemotactic gradients through spatial sensing mechanisms.  This bound is independent of diffusion rate and concentration of the chemoattracts and is instead set  by detachment rate of ligands  from the cell receptors. The result could help explain experimental observations.
\end{abstract}

\maketitle

 \section*{Introduction}
 Cells have to make sense out of their surroundings. They have to go towards sources of nutrition and away from sources of danger.  Berg and Purcell \cite{berg},  found that     concentration measurement by  cells such as  \emph{E. Coli} approach  one of    optimum design. The fluctuations in ligand receptor binding were later incorporated   by \cite{bialek} who found that cells perform concentration measurements within limits set by these fluctuations. Since then, there have been many theoretical  works in understanding the  limitations imposed on concentration measurements by the cell. Some of these include \cite{szabo} that considered effect of ligand diffusion on fluctuations in occupancy of receptors. Understanding how increasing number of receptors could affect reduction in  measurement noise due to receptor noise were considered in \cite{rappel}.  Corrections to contributions to diffusive arrival of ligands obtained by \cite{bialek} were considered in \cite{ten wolde}. Constraints placed by energy consumption in concentration measurement of cells was considered in \cite{mehta}-\cite{energy2}. Limits to concentration sensing by a cell in an environment of interfering ligands was considered in \cite{mora}. Maximum likelihood estimation of concentration of ligands by looking at history of attachment detachment of ligands to the receptor was considered in \cite{wingreen endres 3}. Most calculations that involve evaluation of limits to concentration detection assume that  the cell receptors have reached   equilibrium with the surrounding ligands after which the  attachment/detachment dynamics is considered. However realistic calculations to error measurements should consider measurement of concentration beginning the    moment exposure to chemoattractants occurs.   We label this scenario as the non-equilibrium case in comparison to the equilibrium case which was considered in the  Berg Purcell calculation where the receptor is assumed to have been in equilibrium with the ligand concentration throughout the measurement history.   In the first part of the paper we   show that the bound on error in concentration measurement in the non-equilibrium scenario is  still given by    what is expected from  Berg-Purcell like studies, in the limit of large measurement times. Berg and Purcell considered the quantity getting measured to be the fraction of time the receptor is occupied by the ligand.On  considering  a generic   quantity defining a measurement as the linear combination      of powers of the fraction of time a  receptor is occupied, we show that the error in concentration measurement is similar in the equilibrium and non-equilibrium cases for large measurement times.
 
 
  In addition to concentration measurement, cells also have to measure extracellular gradients of chemoattractants. Cells have evolved to measure gradients with great accuracy. Cells can detect gradients of $1-2 \%$ difference across the cell \cite{gradient exp1}- \cite{gradient exp3}. An optimal response has been seen in the cells where the difference in receptor occupancy between front and back of cells is only $10$ occupied receptors \cite{gradient exp4}.  Understanding the limits to chemotactic gradient measurement by the cells were studied in \cite{wingreen endres} which idealized the cell as a perfectly absorbing sphere and a perfectly monitoring sphere . Fluctuation dissipation theorem was used to consider limitations imposed on gradient sensing  due to ligand receptor kinetics in  \cite{wingreen endres 2}.  Modelling the surface of the cell as an Ising spin chain showed improved ability to detect gradients if receptor cooperativity was introduced \cite{levine}. A signal transduction modelling to understand gradient readouts was considered in \cite{levine rappel}. If one considers the works \cite{wingreen endres}, \cite{wingreen endres 2}, it was shown that the limits to gradient measurements by the cell   went like  $ \frac{1}{ \sqrt{D a c T}}$, upto multiplicative constants.  This implies that one could decrease the   time of measurement $T$ to an arbitrary degree,  by increasing the concentration of chemoattractants and still be able to discern the gradient. In the present work we  represent the cell as sphere covered with receptors with a point source of chemoattractants a particular distance away. We  show that there is a bound on the measurement time, below which it is not possible to discern the concentration gradient set up by the source. This   time being independent of the concentration and diffusion rates of the chemoattractant, implies one cannot simply increase these parameters to decrease the measurement times to being as small as possible. For \emph{Dictyostelium} we evaluate this time to be  around $\sim 4 s$. The time duration of the pulses not being sufficient to discern gradients,  could be a reason as to why \emph{Dictyostelium} cells subjected  to $5 s$   pulses \cite{cyclic} do not respond to the chemotactic gradients of varied concentrations,   while  being placed at varied  distances  from the   chemoattractant source.
  
   In the next section we present the calculation which evaluates the   error in concentration  measurement by cells, when the concentration  detection starts as soon as the cells are bought in contact with chemoattractants. The calculation of the minimum time below which it is not possible to discern   concentration gradients is presented in the following section. We end with conclusions.

 \section*{ Concentration measurements commencing after immediate exposure. }
 
\begin{figure}[h!]
 	\includegraphics[scale=.5]{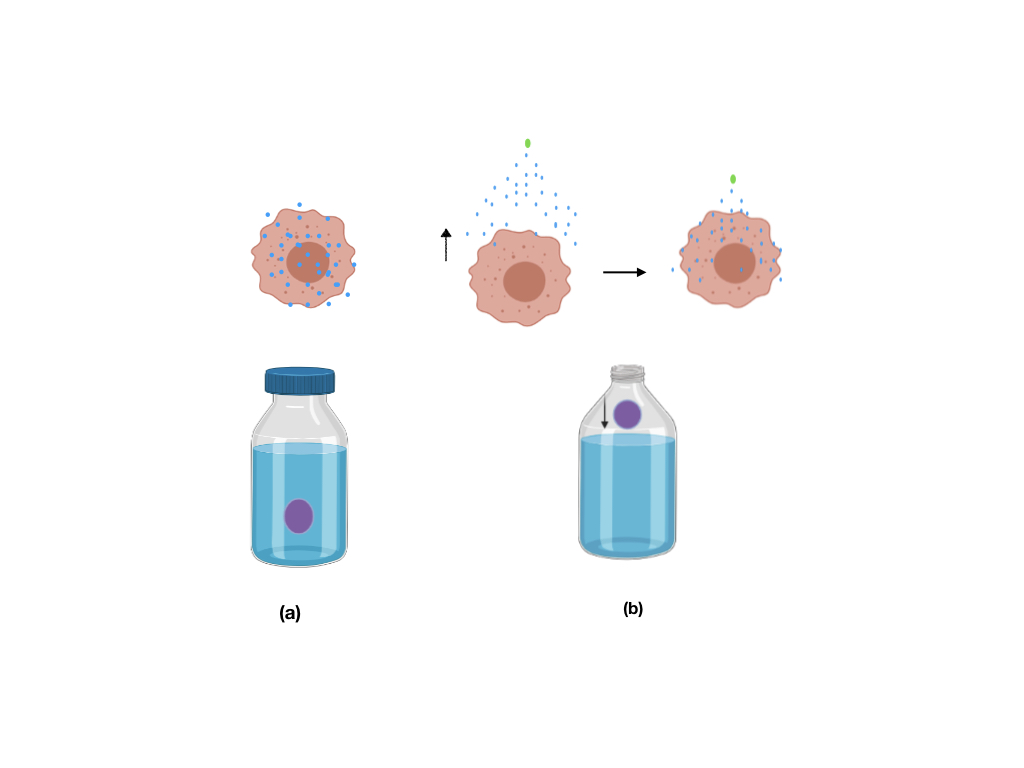}
 	\caption{ In (a) we have a cell in a constant concentration of chemoattractants. The limitations to concentration measurement were studied in the classic work of  Berg Purcell, where  the cell was idealised as a sphere in a constant concentration of chemoattractants. In (b) The cell moves towards a higher concentration of chemoattractants. We  simulate this scenario with the sphere dropped into a bottle of chemoattractants. In this paper we show that the error in concentration measurement in (b) is similar to     the error in   in (a) in limit of large measurement times.   (Images created with BioRender.com) }
 	\label{dropped}
 \end{figure}

 Cells would in general start measuring chemotactic concentration the moment they come in touch with the chemoattractants. Berg and Purcell \cite{berg} assumed the receptor has already equilibrated with the surrounding chemotactic concentration before concentration measurement are considered. Let us instead consider the more realistic case when the cell has started measuring the moment it got in touch with the chemotactic concentration. To understand this, assume the cell represented by a sphere is dropped into a chemoattractant concentration $c$ as shown in Fig. \ref{dropped}. The occupation probability of each cell receptor  obeys the relation 
 \begin{eqnarray}
 \frac{dp( t)}{dt}=k_+ c (1- p( t))-	k_- p( t).
 \end{eqnarray}
 With the initial condition  $p(0)=0$, we get the solution
 \begin{eqnarray}
 p(t)= \frac{k_+c}{k_+ c + k_-} (1-e^{-(k_+c + k_-)t})
 \label{Eq2}
 \end{eqnarray}
 
 Let us say that the receptor  occupancy at time $t$ is given by $O(t)$, which equals to zero if the receptor is not occupied and $1$ if its occupied. The fraction of  time   a receptor is occupied in a time duration $T$ is
 \begin{eqnarray}
 m(T) =\frac{1}{T}\int_0^T O(t)dt
 \end{eqnarray}
 This is something that   the cell can measure.  If we split the time $T$ in to  $\frac{T}{dt}$ intervals, the probability of having $n$ intervals in which receptor was occupied is simply
 \begin{eqnarray}
 e^{-\int_0^T p(t)dt} \frac{[\int_0^T p(t)dt]^n}{n!}
 \end{eqnarray} 
 Hence an average estimate of $m(T)$ would be
 \begin{eqnarray}
 \langle m(T) \rangle &&=\frac{1}{T} \sum_{n=0,\infty} n  e^{-\int_0^T p(t)dt} \frac{[\int_0^T p(t)dt]^n}{n!} = \frac{1}{T} \int_0^T p(t)dt\nonumber\\
  &&= \frac{k_+c}{k_+ c + k_-}-\frac{(1-e^{-(k_+c + k_-)T})}{(k_+ c + k_-)T}\frac{k_+c}{k_+ c + k_-}
  \label{rate of equilibriation}
 \end{eqnarray}
 which in limit of large $T$ becomes
  \begin{eqnarray}
 \langle m(T) \rangle 
 &&= \frac{k_+c}{k_+ c + k_-}\left[ 1-\frac{1}{(k_+ c + k_-)T} \right]\nonumber\\
 &\sim& \frac{k_+c}{k_+ c + k_-}, \quad T \rightarrow \infty
 \label{mean}
 \end{eqnarray}
 The average occupancy   could be inverted by the cell to evaluate the chemoattractant concentration $c$. To evaluate the noise in this estimate of $c$ we need to   evaluate $\langle m(T)^2 \rangle $ for which we need to evaluate $G(t,t')=\langle O(t) O(t') \rangle $.
 Assume the receptor is occupied at time $t$. The  receptor would be still found occupied  at time $t'$, if the receptor remains attached for some time $\tau$ after time $t$, then has $n$ pairs of time intervals     $ t_d^i, t_a^i$ with $i\in [0,n]$, where the receptor is in a detached/ attached state, such that $ \tau + \sum_{i=1,n} t_d^i + t_a^i = t-t'$. For any particular timing combination, the probability of realising the  same   is
 
 \begin{eqnarray}
 &&e^{-\int_0^\tau k_- dx} [k_+c k_-]^n\Pi_{i=1,n} dt_d^i dt_a^i  e^{ -k_+ c\int_0^{t_d^i} dx} e^{ -k_-\int_0^{t_a^i} dx} \nonumber\\
 \end{eqnarray}
 
 One then sums over all possible timing combinations and all values of $n$ to get the probability of still finding the receptor occupied  at time $t'$. Let us call this probability $P(t,t')$.  Since the only information getting into the evaluation of $P(t,t')$ is that the ligand is attached at time $t$ and since the ligand attachment detachment is  Markovian,   the past history before attachment at time $t$ is irrelevant, so whether the receptor has equilibrated with the ligands or not is irrelevant in evaluation of $P(t,t')$. It is obvious that $G(t,t')=p(t)P(t,t') \quad t< t'$.  Since, we already know from \cite{berg} that in the case where   receptor  has   equilibrated with the chemoattractant concentration 
 \begin{eqnarray}
 G_{eq}(t,t')= \bar{p}^2 +\bar{p}(1-\bar{p} )e^{-\frac{k_-|t-t'|}{1-\bar{p}}}
 \end{eqnarray}
 with
 \begin{eqnarray}
 \bar{p} = \frac{k_+c}{k_+ c + k_-}
 \end{eqnarray}
 We get  
 \begin{eqnarray}
 P(t,t')= \bar{p} +(1-\bar{p} )e^{-\frac{k_-|t-t'|}{1-\bar{p}}}
 \label{Eq P}
 \end{eqnarray}
So,
 \begin{eqnarray}
 G(t,t')=p(t) [  \frac{k_+c}{k_+ c + k_-}   +  \frac{k_-}{k_+ c + k_-}e^{- (k_+c + k_-)|t'-t| }], \quad t< t'
 \label{G formula}
 \end{eqnarray}
 Hence,
 \begin{eqnarray}
  &&\langle m(T)^2 \rangle \nonumber\\
 &=&  \frac{1}{T^2}\int_0^T  \int_0^T G(t,t') dt dt'\nonumber\\
 &=&   \frac{1}{T^2}\int_0^T dt \int_0^t dt'     p(t') P(t',t)\nonumber\\ 
 &+&   \frac{1}{T^2}\int_0^T dt' \int_0^{t'} dt     p(t)  P(t,t')   \nonumber\\ 
 &=&   \frac{2}{T^2}\int_0^T dt' \int_0^{t'} dt     p(t)   P(t,t')   \nonumber\\ 
 \label{msquared}
 \end{eqnarray}
 
 The second last equation arises from the one above, because $P(t,t') = P(t',t)$ and on interchanging $t'\leftrightarrow t$,  we find the two integrals in the sum are equal.   Hence as shown in the Appendix, in the limit  where $T$ is very large ($T^{-1}<< k_+ c, k_-$). 
  
 \begin{eqnarray}
 \langle m(T)^2 \rangle - \langle m(T) \rangle^2
 &=&  \left( \frac{ k_+c }{k_+ c + k_-}\right)^2\frac{ 2 }{ (k_+ c + k_-)T} \frac{k_-}{k_+c}
 \label{variance_calculation}
 \end{eqnarray}
 and
  \begin{eqnarray}
  \frac{\delta c}{c} 
 &=& \sqrt{   \frac{ 2(k_+ c + k_-)}{  T} \left[ \frac{k_-}{k_+c}\right]}\frac{1}{    k_- }.\nonumber\\
 \label{equilibrium}
\end{eqnarray}
  
 This is the same result as obtained by Berg and Purcell \cite{berg}. What we hence see is that the error in the non-equilibrium case   is the same as the equilibrium case when $(k_+ c + k_-)T>>1$. Berg and Purcell assumed that the measurement made by the cell was the fraction of time the receptor was occupied. Receptor activation starts a series of downstream reactions in the cell that leads to a cellular response. There is no reason to expect that the cellular response however it is quantified would simply be proportional to the fraction of time the receptor was occupied. It is more likely that as a general case    the measurement   could be proportional to a linear combinations of powers of the fraction of time the receptor was occupied.  Such a quantity could be written as  $Q(T) = \sum_{i>0} a_i m(T)^i$, where $a_i$ are constants dependent on the nature of the system being analysed and the corresponding readout.  Now,

%

  \begin{eqnarray}
  \langle m(T) \rangle^n &&= [ \frac{1}{T}\int_0^T p(t) dt]^n = [\frac{k_+c}{k_+ c + k_-} ]^n[ 1- \frac{1}{T}\int_0^T e^{-(k_+c + k_-)t} dt]^n\nonumber\\
  &&= [\frac{k_+c}{k_+ c + k_-} ]^n[ 1- \frac{n}{T}\int_0^T e^{-(k_+c + k_-)t} dt + \frac{n(n-1)}{2! T^2}[\int_0^T e^{-(k_+c + k_-)t}  dt]^2]  \nonumber\\
 &&=\langle m(T) \rangle_{eq}^n  -[\frac{k_+c}{k_+ c + k_-} ]^n\frac{n}{(k_+ c + k_-)T}   + \mathcal{O}(\frac{1}{T^2}),  \quad if \; (k_+c + k_-)T>>1
 \end{eqnarray}
 
  Also,
  
  \begin{eqnarray}
 &&\langle m(T)^n \rangle = \frac{1}{T^n}\int_0^T dt_1...\int_0^T dt_n \langle O(t_1)...O(t_n) \rangle\nonumber\\
 &=&   \frac{n!}{T^n}\int_0^T dt_{n-1}... \int_0^{t_1} dt     p(t)   P(t,t_1) P(t_1,t_2)...  P(t_{n-2},t_{n-1})  \nonumber\\ 
&=&\langle m(T)^n \rangle_{eq}   -    \frac{n!}{T^n}\int_0^T dt_{n-1}... \int_0^{t_1} dt  \frac{k_+c}{k_+ c + k_-}  e^{-(k_+c + k_-)t}     [  \frac{k_+c}{k_+ c + k_-}   +  \frac{k_-}{k_+ c + k_-}e^{- (k_+c + k_-)|t_1-t| }]..\nonumber\\
&&..[  \frac{k_+c}{k_+ c + k_-}   +  \frac{k_-}{k_+ c + k_-}e^{- (k_+c + k_-)|t_{n-1}-t_{n-2}| }]  \nonumber\\
&=&\langle m(T)^n \rangle_{eq}   -  [  \frac{k_+c}{k_+ c + k_-}]^n \frac{n!}{T^n}\int_0^T dt_{n-1}... \int_0^{t_1} dt    e^{-(k_+c + k_-)t}     \nonumber\\
&& -\frac{k_-}{k_+ c + k_-} [  \frac{k_+c}{k_+ c + k_-}]^{n-1}   \frac{n!}{T^n}\int_0^T dt_{n-1}... \int_0^{t_1} dt     e^{-(k_+c + k_-)t}[ e^{- (k_+c + k_-)|t_1-t| } +...+ e^{- (k_+c + k_-)|t_{n-1}-t_{n-2}| } ] \nonumber\\
&&+ terms \; higher \; order \; in \; \frac{1}{T} \nonumber\\
&&\sim \langle m(T)^n \rangle_{eq}   -[\frac{k_+c}{k_+ c + k_-} ]^n\frac{n}{(k_+ c + k_-)T} + \mathcal{O}(\frac{1}{T^2}) ; \quad if \; (k_+c + k_-)>>1
 \end{eqnarray}
 The factor of $ n!$ in the second equation has a origin similar to the factor of $2$ in Eq.\ref{msquared}.   Now,
%
  
     
  \begin{eqnarray}
 \langle m(T)^n \rangle_{eq} -  \langle m(T) \rangle_{eq}^n &=&  \frac{n!}{T^n}\int_0^T dt_{n-1}... \int_0^{t_1} dt  \frac{k_+c}{k_+ c + k_-}       [  \frac{k_+c}{k_+ c + k_-}   +  \frac{k_-}{k_+ c + k_-}e^{- (k_+c + k_-)|t_1-t| }]..\nonumber\\
 &&..[  \frac{k_+c}{k_+ c + k_-}   +  \frac{k_-}{k_+ c + k_-}e^{- (k_+c + k_-)|t_{n-1}-t_{n-2}| }]     -[\frac{k_+c}{k_+ c + k_-}]^{n }\nonumber\\
 &=&\frac{k_-}{k_+ c + k_-} [  \frac{k_+c}{k_+ c + k_-}]^{n-1}   \frac{n!}{T^n}\int_0^T dt_{n-1}... \int_0^{t_1} dt     [ e^{- (k_+c + k_-)|t_1-t| } +...+ e^{- (k_+c + k_-)|t_{n-1}-t_{n-2}| } ] \nonumber\\
 &&+ terms \; higher \; order \; in \; \frac{1}{T} \nonumber\\
 &=&\frac{k_-}{k_+ c + k_-} [  \frac{k_+c}{k_+ c + k_-}]^{n-1}   \frac{n(n-1) }{ (k_+ c + k_- )T } + \mathcal{O}(\frac{1}{T^2}),\quad if \; (k_+c+k_-)T>>1  \nonumber\\
  \end{eqnarray}
%
%
%
Hence,
\begin{eqnarray}
&& \langle Q(T)^2 \rangle - \langle Q(T) \rangle^2 \nonumber\\
&=&   \sum_{i,j >0} a_i a_j [ \langle m(T)^{i+j} \rangle -   \langle  m(T)^{i }\rangle\langle  m(T)^{j }\rangle] \nonumber\\
&=&   \sum_{i,j >0} a_i a_j  [  \langle m(T)^{i+j} \rangle_{eq}  -[\frac{k_+c}{k_+ c + k_-} ]^{i+j}\frac{i+j}{(k_+ c + k_-)T} + \mathcal{O}(\frac{1}{T^2})   \nonumber\\
&&-  [ \langle  m(T)^{i }\rangle_{eq}  -[\frac{k_+c}{k_+ c + k_-} ]^{i }\frac{i }{(k_+ c + k_-)T} + \mathcal{O}(\frac{1}{T^2})] [\langle  m(T)^{j }\rangle_{eq}   -[\frac{k_+c}{k_+ c + k_-} ]^{ j}\frac{ j}{(k_+ c + k_-)T} + \mathcal{O}(\frac{1}{T^2})]  ] \nonumber\\
&& ,\quad if \; (k_+c+k_-)T>>1  \nonumber\\
&=&  \sum_{i,j >0} a_i a_j [ \langle m(T)^{i+j} \rangle_{eq} -   \langle  m(T)^{i }\rangle_{eq}\langle  m(T)^{j }\rangle_{eq} ] + \mathcal{O}(\frac{1}{T^2}) ,\quad if \; (k_+c+k_-)T>>1  \nonumber\\
 &=&\langle Q(T)^2 \rangle_{eq} - \langle Q(T) \rangle_{eq}^2 + \mathcal{O}(\frac{1}{T^2}),\quad if \; (k_+c+k_-)T>>1 \nonumber\\
&=&\langle Q(T)^2 \rangle_{eq} - \langle Q(T) \rangle_{eq}^2 ,\quad if \; (k_+c+k_-)T>>1 \nonumber\\
\end{eqnarray}
 This is because $  \sum_{i,j >0} a_i a_j [ \langle m(T)^{i+j} \rangle_{eq} -   \langle  m(T)^{i }\rangle_{eq}\langle  m(T)^{j }\rangle_{eq} ]$  is $ \mathcal{O}(\frac{1}{T }) \quad if \; (k_+c+k_-)T>>1$, as it contains terms of the form $ \langle m(T)^n \rangle_{eq} -  \langle m(T) \rangle_{eq}^n$ which are $ \mathcal{O}(\frac{1}{T })$ in the same limit. For a specially chosen combination of $\{ a_i\}$'s, we could have that $ \sum_{i,j >0} a_i a_j [ \langle m(T)^{i+j} \rangle_{eq} -   \langle  m(T)^{i }\rangle_{eq}\langle  m(T)^{j }\rangle_{eq}] $ may be of $\mathcal{O}(\frac{1}{T^2})$, but for a generic combination of $\{ a_i\}$'s, this is not the case and hence $\langle Q(T)^2 \rangle - \langle Q(T) \rangle^2 = \langle Q(T)^2 \rangle_{eq} - \langle Q(T) \rangle_{eq}^2 ,\quad if \; (k_+c+k_-)T>>1 $.  Also,

\begin{eqnarray}
\langle Q(T)  \rangle &=& \sum_{i  >0} a_i    \langle m(T)^{ i} \rangle \nonumber\\
&= & \sum_{i  >0} a_i     [\frac{k_+c}{k_+c + k_-}]^{ i}   + \mathcal{O}(\frac{1}{T}) ,\quad if \; (k_+c+k_-)T>>1\nonumber\\
&= & \langle Q(T)  \rangle_{eq} + \mathcal{O}(\frac{1}{T }) ,\quad if \; (k_+c+k_-)T>>1\nonumber\\
&= & \langle Q(T)  \rangle_{eq} ,\quad if \; (k_+c+k_-)T>>1
\end{eqnarray}

 We hence see that for   generic $\{ a_i\}$'s, the evaluation of $\frac{\delta c}{c}$ is the same in the limit $ (k_+c+k_-)T>>1$, whether we consider the Berg Purcell like equilibrium calculations which assume  that the receptor has already equilibrated with the ligands before measurement starts or we consider the realistic non equilibrium case where the measurement commences the moment the receptor is bought in contact with the ligands.

 \section*{Lower bound on gradient measurement times in cells.}

\begin{figure}[h]
		\includegraphics[scale=.5]{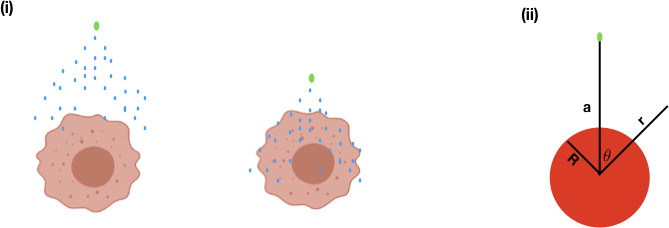}
		\caption{ (i) A cell   getting exposed to a point source of chemoattractants.  Experiments have shown   \cite{cyclic}, that for stimulation with $5 s$ pulses  for a range of cAMP concentrations, no evidence of chemotaxis was observed at any distance from the cAMP pipette. Calculations in this paper suggest a measurement time bound of $4 s$, below which it is not possible for cells to discern gradients irrespective of concentration of chemoattractants involved. (Images reated with BioRender.com.)  (ii) The various quantities entering into the calculation in the text. The cell is represented by a sphere of radius $R$, and the source of chemoattractant is represented by in green colour at a distance of $a$ from the center of the sphere.   }
		\label{figure}
	\end{figure}

	To consider the problem of time required to measure a chemotactic gradient by the cell, consider the following setup illustrated in Fig.\ref{figure}. A sphere representing the cell is covered with receptors, such that the probability of the receptor at   position $(R,\theta)$ being occupied is   $p(\theta,t)$ which   at equilibrium   becomes $p(\theta)$. The source of chemottractants is at a distance $a$ from the center of the sphere. Let us say  the concentration of chemoattractants at a position $(r,\theta)$, is $c(r,\theta,t)$ which at equilibrium becomes $c(r,\theta)$. We  hence have that 
		\begin{eqnarray}
\frac{dp(\theta,t)}{dt}=k_+ c(R,\theta,t)(1- p(\theta,t))-	k_- p(\theta,t)
	\end{eqnarray}
	and
	\begin{eqnarray}
-D \frac{\partial c(r,\theta,t)}{\partial r} |_{r=R}= k_+ c(R,\theta,t)(1- p(\theta,t))-	k_- p(\theta,t)
	\end{eqnarray}
		 At equilibrium we   have
	\begin{eqnarray}
 k_- p(R,\theta)&&= k_+ c(R,\theta)(1- p(R,\theta))\nonumber\\
 D \frac{\partial c(r,\theta)}{\partial r} |_{r=R}&&=0
	\end{eqnarray}
	The symmetry of the problem implies no $\phi$ dependence. 	Since the source of chemoattractants is located at $r=a$, we have the equation
	\begin{eqnarray}
D \nabla^2 c(r,\theta,\phi) = C \delta(r-a)\delta(\theta)
\label{source}
	\end{eqnarray}
 whose solution as described in the Appendix is 
 \begin{eqnarray}
	c(R,\theta)  
	&=& \frac{C}{4\pi D a} + \frac{C}{4\pi D a}\sum_{n=1,\infty}  \frac{R^n}{a^{n}} P_n(\cos \theta).
 \end{eqnarray}
Since
 	\begin{eqnarray}
  p(R,\theta)&&= \frac{k_+ c(R,\theta)}{ k_+ c(R,\theta) + k_-},\nonumber\\
 \end{eqnarray}
  we have the error in concentration measurement  \cite{berg} for large times 
 \begin{eqnarray}
\frac{\triangle c_{rms} (R,\theta)}{c(R,\theta)} &&=\sqrt{ \frac{2}{k_- T p(R,\theta)}   }=\sqrt{ \frac{2(k_+c(R,
		\theta) + k_-))}{k_-T k_+c    }   } \nonumber\\
 \end{eqnarray}
 However, including terms to all orders in $\frac{1}{T}$, we have.
 
 \begin{eqnarray}
 \frac{\triangle c_{rms} (R,\theta)}{c(R,\theta)} &&=  \sqrt{ \frac{2  [(k_+ c(R,\theta ) T + k_-T ) - (1-e^{-(k_+ c(R,\theta ) + k_-)T })] }{(k_- T)( k_+ c(R,\theta )  T  ) }  }\nonumber\\
 \end{eqnarray}
 Hence, in order to be able to discern a gradient  we should have that
 \begin{eqnarray}
c(R,\theta = 0)  - 	c(R,\theta = \pi)  &>& \triangle c_{rms}(R,\theta = 0)  +	\triangle c_{rms}(R,\theta = \pi) \nonumber\\
&=& \sqrt{ \frac{2c(R,\theta = 0) [(k_+ c(R,\theta = 0) T + (k_-T  - 1) + e^{-(k_+ c(R,\theta = 0) + k_-)T })] }{k_- k_+ T^2  }  }\nonumber\\
&+& \sqrt{ \frac{2c(R,\theta = 0) [(k_+ c(R,\theta = \pi) T + (k_-T  - 1) + e^{-(k_+ c(R,\theta = \pi) + k_-)T })] }{k_- k_+ T^2  }  }\nonumber\\
\end{eqnarray}
 
Assume that the measurement time $T$ is such that $k_-T>1$, then
\begin{eqnarray}
c(R,\theta = 0)  - 	c(R,\theta = \pi) &>& \sqrt{ \frac{2c(R,\theta = 0) (k_+ c(R,\theta = 0)    )}{k_- k_+ T  }  }\nonumber\\
&+& \sqrt{ \frac{2c(R,\theta = \pi ) (k_+ c(R,\theta = \pi)  )}{k_- k_+ T  }}   \nonumber\\
 \frac{C}{4\pi D a} \frac{2R/a}{1-R^2/a^2}&>&\sqrt{ \frac{2 }{k_-   T  }}[ c(R,\theta = 0)  + 	c(R,\theta = \pi)]\nonumber\\
 \frac{C}{2\pi D a}\frac{R/a}{1-R^2/a^2}&>&\sqrt{ \frac{2 }{k_-   T  }}[   \frac{C}{2\pi D a} + \frac{C}{2\pi D a}\frac{R^2/a^2}{1-R^2/a^2}   ]\nonumber\\
T&>& \frac{2}{k_-}[\frac{  1  +\frac{R^2/a^2}{1-R^2/a^2}   }{  \frac{R/a}{1-R^2/a^2}}]^2 =\frac{2}{k_-}\frac{a^2}{R^2}
 \end{eqnarray}
 We hence get a bound on the measurement time, independent of the nature of the chemoattractant or its concentration, below which it is not possible for the cell to decipher the concentration gradient.   The smallest value this bound can take is when $a=R$ or $T_{smallest} =\frac{2}{k_-}$. Since  $k_-T\geq 2$, the assumption $k_-T\geq 1$ used to derive   above implies consistency of the derivation above. 
 Experiments  \cite{rates} have evaluated the detachment rates of Cy3-cAMP molecules from receptors on the surface of dictyostelium and found that $k_{-}$ was in the range of $1.1-.39 s^{-1}$ in the anterior pseudopod region of the cell and $.39-.1 s^{-1}$ in the posterior tail region.  So assuming an rate of  $k_{-} \sim .5 s^{-1}$,   gives a bound on $T_{smallest} =\frac{2}{k_-} = 4 s$. This is   in line with   \cite{cyclic}   where it was seen that the  simulation of \emph{Dictyostelium} with $5 s$ pulses using a wide range of cAMP concentrations, shows no evidence of chemotaxis by cells at any distance from the pipette. A possible reason for this could be the fact that the time period for which the signal was active,  was not sufficient for discerning the gradient by the cell, leading to no response.

 \section*{Conclusion}
 Any calculation that evaluates the error in measurement of concentation should consider the cell from the time when it first gets in touch with the ligand concentration. In the first part of the paper we evaluated the error in such a scenario. We showed that the error is the same, in the limit of large measurement time, as evaluated in the Berg Purcell calculation   which assumed that the cell was already in equilibrium  with the ligand concentration when measurement commenced.  In the non-equilibrium case and for large measurement times $ T>> (k_+c + k_-)^{-1}$, the receptor will have   equilibrated with the ligand concentration for most of  its measurement history and hence it is expected that   $ \langle m(T) \rangle$ as well  $\langle m(T)^2 \rangle$ will yield   result similar to   the case in which the receptor was already in equilibrium with the surrounding concentration when the measurement commenced. However, there is no reason apriori that the difference between the two  $\langle m(T)^2 \rangle - \langle  m(T) \rangle^2 $ will yield similar values. We however see an equality between the equilibrium and non-equilibrium estimates. The cellular response would generally   be proportional to   some linear combination of powers of the fraction of time the receptor is  occupied. By considering  the      measured quantity  to go as  $Q(T) = \sum_{i>0} a_i m(T)^i$, where the $\{a_i\}$'s are generically chosen, we found that the error in concentration measurement in the equilibrium as well as non equilibrium cases is the same for large measurement times.

  \cite{bialek} have  utilised a fluctuation dissipation framework to talk about how noisy ligand  attachment/detachment events add limitations to concentration measurements. Their work first considers the case without diffusion, producing an rms error in the estimate of receptor occupancy $n$ given by
  \begin{eqnarray}
  \delta n_{rms} =\sqrt{ \frac{2\bar{n}(1-\bar{n})}{(k_+ \bar{c} + k_-)\tau}}
  \end{eqnarray}
  including diffusion increases this error to
  \begin{eqnarray}
  \delta n_{rms} =\sqrt{ \frac{2\bar{n}(1-\bar{n})}{(k_+\bar{c} + k_-)T} +\frac{(
  		\bar{n}(1-\bar{n}))^2}{\pi D a \bar{c}T}}
  \end{eqnarray}
  where $\bar{c},\bar{n}$ are averages of measured concentration and receptor occupancy, $T$ is the measurement time, $D$ is the diffusion coefficient and $a$ is the receptor size. Hence, including the effect of diffusion only serves to increase the error. Since  \cite{bialek}, considered the   situation where the ligands have equilibrated with the receptor, they could use the fluctuation dissipation frameworks to do their calculations. Our work in part one of the paper is considering the case when the receptor is suddenly got in touch with the ligands implying a non-equilibrium framework, so fluctuation dissipation frameworks to do calculations are not possible here. Hence, including diffusion effects if possible would be more involved and we do not attempt the same in our paper. However from the lessons of equilibrium calculations as done in \cite{bialek}, we can say that diffusion effects will only add to measurement errors. How this addition will look like or whether it will yield the same form as equation above, given the similarity in the equilibrium and non-equilibrium case when    diffusion is not considered, is an open question.
  
   It is also quite interesting   that  the observation that the  bound on   measurement times below which gradients cannot be detected,  being independent of the chemoattractant properties such as concentration and diffusion  was not noticed in theoretical works up until now.  For example in \cite{wingreen endres 2} the authors consider the noise in receptor ligand  binding/unbinding to evaluate the limits to detection of gradients.  The only information of the chemotactic gradient that appears in their analysis is the ligand concentration  at points on the surface of the cell. For example   the section \emph{Two Receptors} evaluates the  variance in the difference between concentration between two points on the cell surface and relates the same to the concentration at these two points.  The work does not calculate any relation between  the ligand concentration of various points on the cell surface, but assumes the same to be given. In our work we consider a point source that produces the  chemotactic gradient around the cell itself and calculate the concentration on various points on the cell surface. Using this crucial information we could evaluate a bound on the measurement time itself   below which no gradient could be discerned irrespective of concentrations involved, an observation that was missed/not obtained     in works such as \cite{wingreen endres 2}.

 The FRET measurements for example in \cite{fret}  point to various signalling elements in the cell getting activated, however they do not imply or suggest that any chemotaxis in direction of the source is attained for a pulse of magnitude below 5 seconds, since as is known from \cite{cyclic} there is no chemotactic response in direction of the gradient for pulses of time duration 5 sec. The part two of the paper provides a reason as to why for all possible ligand     concentrations     there is no chemotactic response observed for the 5 second pulse, tracing this reason  to the limitations imposed by the ligand binding unbinding noise. 

  \section*{Appendix}
 
   \subsection*{ I}
   We elaborate on some calculations from the section on concentration measurements. We know that
  \begin{eqnarray}
   &&\langle m(T)^2 \rangle \nonumber\\
   &=&  \frac{1}{T^2}\int_0^T  \int_0^T G(t,t') dt dt'\nonumber\\
   &=&   \frac{1}{T^2}\int_0^T dt \int_0^t dt'     p(t') [  \frac{k_+c}{k_+ c + k_-}   +  \frac{k_-}{k_+ c + k_-}e^{- (k_+c + k_-)(t-t') }]   '\nonumber\\ 
    &+&   \frac{1}{T^2}\int_0^T dt' \int_0^{t'} dt     p(t) [  \frac{k_+c}{k_+ c + k_-}   +  \frac{k_-}{k_+ c + k_-}e^{- (k_+c + k_-)(t'-t) }]   \nonumber\\ 
      &=&   \frac{2}{T^2}\int_0^T dt' \int_0^{t'} dt     p(t) [  \frac{k_+c}{k_+ c + k_-}   +  \frac{k_-}{k_+ c + k_-}e^{- (k_+c + k_-)(t'-t) }]  \nonumber\\ 
       &=&   \frac{2}{T^2}\int_0^T dt' \int_0^{t'} dt     [ \frac{k_+c}{k_+ c + k_-} ](1-e^{-(k_+c + k_-)t}) [  \frac{k_+c}{k_+ c + k_-}   +  \frac{k_-}{k_+ c + k_-}e^{- (k_+c + k_-)(t'-t) }]  \nonumber\\ 
       &=& \frac{1}{T^2}\left( -\frac{2 c k_- k_+}{\left(c k_++k_-\right){}^4}+\frac{2 c k_- k_+ e^{-T \left(c k_++k_-\right)}}{\left(c k_++k_-\right){}^4}+\frac{c^2 k_+^2}{\left(c k_++k_-\right){}^4}-\frac{c^2 k_+^2 e^{-T \left(c k_++k_-\right)}}{\left(c k_++k_-\right){}^4}\right)  \nonumber\\
       & + &  \frac{1}{T} \left(-\frac{c^3 k_+^3}{\left(c k_++k_-\right){}^4}+\frac{c^2 k_- k_+^2 e^{-T \left(c k_++k_-\right)}}{\left(c k_++k_-\right){}^4}+\frac{c k_-^2 k_+ e^{-T \left(c k_++k_-\right)}}{\left(c k_++k_-\right){}^4}+\frac{c k_-^2 k_+}{\left(c k_++k_-\right){}^4}\right) \nonumber\\
       & + &  \left(\frac{c^4 k_+^4}{2 \left(c k_++k_-\right){}^4}+\frac{c^3 k_- k_+^3}{\left(c k_++k_-\right){}^4}+\frac{c^2 k_-^2 k_+^2}{2 \left(c k_++k_-\right){}^4}\right)
   \end{eqnarray}
   where we plugged in to the second last equation Eq.\ref{Eq2}
   
   Also we have 
   \begin{eqnarray}
     &&\langle m(T)  \rangle^2 \nonumber\\
     &=&[ \int_0^T  \frac{k_+c}{k_+ c + k_-} ] (1-e^{-(k_+c + k_-)t})]^2\nonumber\\
     &=&  \frac{1}{T^2}\left(\frac{c^2 k_+^2}{\left(c k_++k_-\right){}^4}+\frac{c^2 k_+^2 e^{-2 T \left(c k_++k_-\right)}}{\left(c k_++k_-\right){}^4}-\frac{2 c^2 k_+^2 e^{-T \left(c k_++k_-\right)}}{\left(c k_++k_-\right){}^4}\right)\nonumber\\
     &+ &   \frac{1}{T}\left(\frac{2 c^3 k_+^3 e^{-T \left(c k_++k_-\right)}}{\left(c k_++k_-\right){}^4}-\frac{2 c^3 k_+^3}{\left(c k_++k_-\right){}^4}+\frac{2 c^2 k_- k_+^2 e^{-T \left(c k_++k_-\right)}}{\left(c k_++k_-\right){}^4}-\frac{2 c^2 k_- k_+^2}{\left(c k_++k_-\right){}^4}\right)\nonumber\\
     &+&  \left(\frac{c^4 k_+^4}{\left(c k_++k_-\right){}^4}+\frac{2 c^3 k_- k_+^3}{\left(c k_++k_-\right){}^4}+\frac{c^2 k_-^2 k_+^2}{\left(c k_++k_-\right){}^4}\right)
   \end{eqnarray}

  Hence
    \begin{eqnarray}
   &&\langle m(T)^2 \rangle - \langle m(T) \rangle^2  \nonumber\\
   &=&\frac{1}{T^2}\left( -\frac{c^2 k_+^2 e^{-2 T \left(c k_++k_-\right)}}{\left(c k_++k_-\right){}^4}+\frac{c^2 k_+^2}{\left(c k_++k_-\right){}^4}+\frac{4 c k_- k_+ e^{-T \left(c k_++k_-\right)}}{\left(c k_++k_-\right){}^4}-\frac{4 c k_- k_+}{\left(c k_++k_-\right){}^4} \right)\nonumber\\
    &+& \frac{1}{T } \left(  -\frac{2 c^3 k_+^3 e^{-T \left(c k_++k_-\right)}}{\left(c k_++k_-\right){}^4}+\frac{2 c^2 k_- k_+^2}{\left(c k_++k_-\right){}^4}+\frac{2 c k_-^2 k_+ e^{-T \left(c k_++k_-\right)}}{\left(c k_++k_-\right){}^4}+\frac{2 c k_-^2 k_+}{\left(c k_++k_-\right){}^4} \right)
   \end{eqnarray}
  
 which in large $T$ limit is 
 \begin{eqnarray}
 \langle m(T)^2 \rangle - \langle m(T) \rangle^2
 &=&  \left( \frac{ k_+c }{k_+ c + k_-}\right)^2\frac{ 2 }{ (k_+ c + k_-)T} \frac{k_-}{k_+c}
 \end{eqnarray}
  
  Since in limit of large $T$ ($T^{-1}<<k_+c, k_-$)
  \begin{eqnarray}
  1 &=&\frac{k_-}{k_+ c}\frac{\langle m(T) \rangle}{1- \langle m(T) \rangle}\nonumber\\
  \frac{\delta c}{c} &=&  \frac{\delta \langle m(T) \rangle}{\langle m(T) \rangle(1- \langle m(T) \rangle)}\nonumber\\
  &=& \sqrt{  \left( \frac{ k_+c }{k_+ c + k_-}\right)^2 \frac{ 1}{ (k_+ c + k_-)T} \left[ 2 \frac{k_-}{k_+c} \right]}\frac{(k_+ c + k_-)^2}{ k_+c  k_- }\nonumber\\
  &=&   \sqrt{ \frac{2}{k_+c T } + \frac{2}{k_- T}}\nonumber\\
  \end{eqnarray}

  \section*{II}
  We next, consider calculations involved in evaluating the bound on gradient measuring time. We have to solve 
  \begin{eqnarray}
  D \nabla^2 c(r,\theta,\phi) = C \delta(r-a)\delta(\theta)
  \label{source}
  \end{eqnarray}
   with the boundary condition
  \begin{eqnarray}
  D \frac{\partial c(r,\theta)}{\partial r} |_{r=R}&&=0
  \end{eqnarray}
  The symmetry of the problem implies no $\phi$ dependence. 	 
  One solution to above equation is
  \begin{eqnarray}
  c(r,\theta) &=& \frac{C}{4\pi D \sqrt{r^2 + a^2 - 2 a r \cos\theta}}\nonumber\\
  \end{eqnarray}
  In the region close to $r=R$, the above becomes
  \begin{eqnarray}
  c(r,\theta) &=& \frac{C}{4\pi Da}\sum_{n=0,\infty} \frac{r^n}{a^{n}} P_n(\cos \theta)
  \end{eqnarray}
 where $P_n(\cos \theta)$ are the Legendre Polynomials. $c(r,\theta) + b(r,\theta)  $ is also a solution of Eq.\ref{source} if
  \begin{eqnarray}
  \nabla^2 b(r,\theta) = 0
  \end{eqnarray}
  Solution of above equation for region $r>R$ is
  \begin{eqnarray}
  b(r,\theta) &=&\sum_{n=0,\infty} \frac{A_n}{r^{n+1}} P_n(\cos \theta)
  \end{eqnarray}
  Hence for the region close to  $r=R$,  a solution is
  \begin{eqnarray}
  c(r,\theta) &=& \frac{C}{4\pi D a}\sum_{n=0,\infty} \frac{r^n}{a^{n}} P_n(\cos \theta)+ \sum_{n=0,\infty} \frac{A_n}{r^{n+1}} P_n(\cos \theta)
  \end{eqnarray}
  The boundary condition
  \begin{eqnarray}
  D \frac{\partial c(r,\theta)}{\partial r} |_{r=R}&&=0
  \end{eqnarray}
  implies 
  \begin{eqnarray}
  0 &=&\frac{C}{4\pi D a}\sum_{n=1,\infty} \frac{nR^{n-1}}{a^{n}} P_n(\cos \theta)- \sum_{n=0,\infty} \frac{(n+1)A_n}{R^{n+2}} P_n(\cos \theta)
  \end{eqnarray}
  Hence $A_0 = 0$ and for $n>0$
  \begin{eqnarray}
  A_n = \frac{C}{4\pi D a}\frac{n R^{2n+1} }{(n+1)a^{n}}
  \end{eqnarray}
  Hence,
  \begin{eqnarray}
  c(R,\theta) &=& \frac{C}{4\pi D a}\sum_{n=0,\infty} \frac{R^n}{a^{n}} P_n(\cos \theta)+ \frac{C}{4\pi D a}\sum_{n=1,\infty}  \frac{n R^{2n+1} }{(n+1)a^{n}}\frac{1}{R^{n+1}} P_n(\cos \theta)\nonumber\\
  &=& \frac{C}{4\pi D a} + \frac{C}{4\pi D a}\sum_{n=1,\infty}  \frac{R^n}{a^{n}} P_n(\cos \theta).
  \end{eqnarray}
  
  So
  \begin{eqnarray}
  c(R,\theta = 0)  
  &=& \frac{C}{4\pi D a} + \frac{C}{4\pi D a}\sum_{n=1,\infty}  \frac{R^n}{a^{n}}  \nonumber\\
  &=& \frac{C}{4\pi D a} + \frac{C}{4\pi D a}\frac{R/a}{1-R/a}    \nonumber\\
  &=& \frac{C}{4\pi D a}\frac{1}{1-R/a}    \nonumber\\
  c(R,\theta = \pi)  
  &=& \frac{C}{4\pi D a} + \frac{C}{4\pi D a}\sum_{n=1,\infty} (-1)^n  \frac{R^n}{a^{n}}  \nonumber\\
  &=& \frac{C}{4\pi D a} - \frac{C}{4\pi D a}\frac{R/a}{1+R/a}     \nonumber\\
    &=& \frac{C}{4\pi D a}\frac{1}{1+R/a}    \nonumber\\
  \end{eqnarray}

\end{document}